\documentclass[10pt,conference]{IEEEtran}
\IEEEoverridecommandlockouts
\usepackage{amsmath,amssymb,amsfonts}
\usepackage{algorithmic}
\usepackage{graphicx}
\usepackage{textcomp}
\usepackage{xcolor}
\usepackage{tikz}

\DeclareMathOperator*{\argmin}{arg\,min} % Jan Hlavacek

\def\BibTeX{{\rm B\kern-.05em{\sc i\kern-.025em b}\kern-.08em
    T\kern-.1667em\lower.7ex\hbox{E}\kern-.125emX}}
\usepackage{longtable}
\usepackage{amsmath}
\usepackage{cancel}
\usepackage{amssymb}
\usepackage{longtable}
\usepackage{multirow}
\usepackage{array}
\usepackage{balance}
\usepackage[labelsep=period]{caption}
\usepackage{setspace}
\usepackage{float}
\usepackage{subfig}
\usepackage{lettrine}
\usepackage[noadjust]{cite}
\usepackage[misc,geometry]{ifsym}
\usepackage{cleveref}
\usepackage[english]{babel}
\usepackage{tikz}
\usepackage[letterpaper, left=0.625in, right=0.625in, bottom=1in, top=0.7in]{geometry}
\usepackage[figurename=Fig.]{caption}
\usepackage{arydshln}

\begin{document}

\title{Towards Quantum Annealing for Multi-user NOMA-based Networks}

\author{Eldar Gabdulsattarov, $^{\#}$Khaled Rabie, $^{\circ}$Xingwang Li, and Galymzhan Nauryzbayev$^{\textrm{~\Letter}}$  
	\\% <-this % stops a space
	\IEEEauthorblockA{School of Engineering and Digital Sciences, Nazarbayev University, Nur-Sultan, Z05H0K4, Kazakhstan\\
	$^{\#}$Department of Engineering, Manchester Metropolitan University, Manchester, M15 6BH, UK\\ 
	$^{\circ}$School of Physics and Electronic Information Engineering, Henan Polytechnic University, Jiaozuo 454000, China \\
	Email: \{eldar.gabdulsattarov, galymzhan.nauryzbayev\}@nu.edu.kz, $^{\#}$k.rabie@mmu.ac.uk, 
	$^{\circ}$lixingwang@hpu.edu.cn
	}
}
 
\maketitle

\begin{abstract}
 Quantum Annealing (QA) uses quantum fluctuations to search for a global minimum of an optimization-type problem 
 %that could potentially solve problems 
 faster than classical computers. To meet the demand for future internet traffic and mitigate the spectrum scarcity, this work presents the QA-aided maximum likelihood (ML) decoder for multi-user non-orthogonal multiple access (NOMA) networks as an alternative to the successive interference cancellation (SIC) method. 
 %The system model considered in the analysis is the $3$-user NOMA scenario under BPSK modulation. 
 The practical system parameters such as channel randomness and possible transmit power levels are taken into account for all individual signals of all involved users. The brute force (BF) and SIC signal detection methods are taken as benchmarks in the analysis. The QA-assisted ML decoder results in the same BER performance as the BF method outperforming the SIC technique, but the execution of QA takes more time than BF and SIC. The parallelization technique can be a potential aid to fasten the execution process. 
 %Furthermore, time taken for accessing the QA machine and for error correcting procedures can be effectively reduced in future less noisy quantum annealers. 
 This will pave the way to fully realize the potential of QA decoders in NOMA systems. 
%QA demonstrated potential enhancement ...
%... taken into account when a maximum likelihood estimation is performed for all ...
\end{abstract}


\section{Introduction}

There has been a drastic rise in the traffic of advanced multimedia applications such as real-time conferencing, online video gaming and virtual reality in recent years. Moreover, according to the Cisco Annual Report, the total number of internet users will reach $5.3$ billion by $2023$ \cite{cisco}. To meet this ever-growing demand on traffic, it is crucial to consider some optimization improvements in the quality of data transmission in terms of the bandwidth, error rate, and delay. One of the actions that can be taken into account is to implement Quantum Annealing (QA) technologies to solve NP-hard optimization problems in wireless communication. 

The authors in \cite{ldpc,leveraging,tabi,kizilirmak} suggested the centralized radio access network (C-RAN) as a promising and cost-effective design architecture for future wireless networks, where multiple base stations (BSs) are interconnected with a centralized data center. The C-RAN's data centers support most of the BSs' signal processing operations. It is envisioned that the QA computers placed at the C-RAN centers can facilitate optimization problems, paving the way for full integration of the presented methodologies in real-world communication systems.

One of the computational complex problems in wireless networks is to obtain an optimal scheduling in multi-user (MU) systems, which can help one effectively employ network resources. The authors in \cite{tdma} introduced the QA approach to find an optimal solution for the time-division multiple access (TDMA) transmission scheduling problem in a wireless sensor network using a tree topology. The results revealed that QA outperforms the other scheduling solutions in terms of the runtime and proximity to the optimal solution.

An essential part of cellular base-band processing that can enhance the quality of communication is the error correction code (ECC), which mitigates the errors in received bit-streams caused by the noise in a wireless environment. However, the error correction code protocols with exceptional performance require significant computational resources for the decoding process at the receiver. For this reason, the implementation of QA computers was proposed in \cite{ldpc} for efficient decoding of low-density parity-check codes in uplink systems. The method achieved a bit error rate (BER) of $ 10^{-8} $ at the signal-to-noise ratio (SNR) of $2.5$-$3.5$ dB lower than the standard algorithm for decoding ECCs on a field-programmable gate array (FPGA). 

% At the same time, it is important to consider minimizing the inter-user interference and complexity of data detection in downlink multi-user systems. Vector perturbation precoding (VPP) is a promising technique that can provide such improvement; however, it is limited in its application to small-scale multiple-input multiple-output (MIMO) due to the fact that to, identify an optimal perturbation in VPP, high computational power is needed. Therefore, in \cite{vpp}, the authors proposed a novel method for processing NP-hard VPP using QA (i.e., QAVP) in MIMO downlink systems. The experimental results demonstrated that the method can reach BER of $10^{{-4}}$ at the SNR of $32$ dB with computational time in order of $\mu$s for a $6 \times 6$ MIMO system with $64$-QAM (quadrature amplitude modulation). Moreover, QAVP has the potential to surpass the performance of such techniques as zero forcing (ZF) and fixed complexity sphere encoder.

Another key aspect in the performance of wireless communication systems is a signal detection. To achieve the full potential of multi-user systems, BSs have to efficiently decode superimposed data streams received from different users. Maximum likelihood (ML) usually ends up with an optimum solution for decoding the received signals, but it is an NP-hard problem. The computational cost of ML rises exponentially with the number of users and data rate \cite{leveraging}. For large-scale networks, the ML decoding process might last expiration of acknowledgement timeout. The authors in \cite{leveraging,tabi,kizilirmak} employed a QA approach to speed up this process. Particularly, \cite{leveraging} presented the QuAMax approach for ML decoding in MU MIMO systems using QA. The performance of the solution was tested on the quantum annealer with 2048 qubits considering different modulation schemes and multiple users. The method achieves the same BER as ZF does, but with significantly lower computation time in the case of binary phase-shift keying (BPSK) and quadrature phase-shift keying (QPSK) modulation types. Moreover, the QA approach outperforms the conventional decoders in terms of the computable size of ML problems. For instance, the work in \cite{leveraging} was extended by \cite{tabi}, where an experiment on the recently released QA hardware with $5640$ qubits was performed. The results revealed that the new computer is able to provide better scaling for embedding the ML MIMO decoding problem. In addition, the study showed that today's QA machines can achieve identical BER as the ML sphere decoder for $4 \times 4$ MIMO using $16$-QAM (quadrature amplitude modulation). 

The current communication technologies support multiple users using orthogonal multiple access (OMA) techniques such as TDMA, frequency- and code-division multiple access (i.e., FDMA and CDMA) schemes. However, all these techniques have certain shortcomings such as precise time synchronization, limited frequency carriers and code length. These drawbacks get more perceptible for a large number of users \cite{noma_tutorial}. Non-orthogonal multiple access (NOMA) is a technology that allows all users in a cell operate in the common time and frequency, that, in turn, improve the spectral efficiency (SE). The successive interference cancellation (SIC) is considered as a promising decoding technique for NOMA that decodes symbol-by-symbol until the desired one is detected. However, the high computational cost, latency and error propagation of SIC make it impractical to implement NOMA in the current wireless communication systems \cite{noma_tutorial}, \cite{survey_noma}. The traditional ML decoder supports high-quality decoding but suffers from high computation time. To address this issue, a NOMA-ML decoder based on QA was proposed in \cite{kizilirmak} for a two-user system model, with a BPSK modulation and constant channel gains.

In contrast to the previous studies, in this work, we consider the NOMA-based uplink network comprising multiple users under practical scenarios and different modulation types. The NOMA-ML decoding problem was firstly reformulated into a QUBO model for multiple modulation schemes including BPSK, QPSK, $16-$QAM and $64-$QAM. Moreover, the generalized QUBO expressions of the proposed system model are derived for the multi-user network under the previously defined modulation types. We have investigated how the number of users in a cell affects the signal decoding performance. Furthermore, the QA-assisted ML decoder was compared with SIC and Brute Force (BF) techniques in terms of BER performance and simulation time. Finally, the influence of the parallelization method on the execution time of the QA machine is examined.

%The paper is organized as follows. The system model and the background concepts of NOMA, ML Detection, QA and QUBO are defined in Section \ref{section:system model}. While Section \ref{section:formulation} illustrates QUBO formulation for NOMA ML signal detection. Further, Section \ref{section:implementation} explains the implementation of the problem on a quantum computer and discusses the obtained simulation results. Lastly, the main findings of the work are summarized in Section \ref{section:conclusion}.

\section{System Model}
\label{section:system model}
\begin{figure}[!t]
	\begin{tikzpicture}
		\coordinate (A) at (-2, 0);
		\coordinate (B) at (-1, 0);
		\coordinate (C) at (-1.5, 1.5);
		%\node [above] at (-1.5, 1.8) {BS};
		
		\coordinate (U1) at (0.5,0);
		\coordinate (U2) at (2,0);  
		\coordinate (U3) at (3.5,0);
		\coordinate (UN) at (5.5,0);
		
		%base station
		\draw (A) -- (B) -- (C) -- cycle;
		\draw (C) -- (-1.5, 1.6);
		\draw (-1.5, 1.6) -- (-1.3, 1.6);
		\draw (-1.3, 1.6) -- (-1.3, 1.8);
		\draw (-1.5, 1.6) -- (-1.7, 1.6);
		\draw (-1.7, 1.6) -- (-1.7, 1.8);
		\node [below] at (-1.5,0) {$0$};
		%users points
		\filldraw [black] (U1) circle [radius = 1pt]
		(-1.5,0) circle [radius = 1pt]
		(U2) circle [radius = 1pt]
		(U3) circle [radius = 1pt]
		(UN) circle [radius = 1pt];
		%User1				 				 
		\node at (0.5, 0.125) [rectangle, draw] (user1) {};
		\node [above] at (user1.north) {$U_1$};
		\node [below] at (user1.south) {$50$};
		%User2			 
		\node at (2,0.125) [rectangle, draw] (user2) {};
		\node [above] at (user2.north) {$U_2$};
		\node [below] at (user2.south) {$\frac{50N}{N-1}$};
		%User3				 
		\node at (3.5,0.125) [rectangle, draw] (user3){};
		\node [above] at (user3.north) {$U_3$};
		\node [below] at (user3.south) {$\frac{50(N+1)}{N-1}$};
		%UserN				 
		\node at (5.5,0.125) [rectangle, draw] (userN){};
		\node [above] at (userN.north) {$U_N$};
		\node [below] at (userN.south) {$100$};
		%line
		\draw [->] (-1.5, 0) -- (6.5,0);
		\node [below] at (6.5,0) {\rm m};
		
		%arrows
		\draw [->] (0.2,0.6) -- (-1, 1.5) node [fill = white, midway]{$h_1$};
		\draw [->, bend right = 5] (1.7,0.6) to  node [fill = white, midway]{$h_2$} (-1, 1.6);
		\draw [->, bend right = 10] (3.2,0.6) to node [fill = white, midway]{$h_3$} (-1, 1.7);
		\draw [->, bend right = 10] (5.2,0.6) to node [fill = white, midway]{$h_N$} (-1, 1.8);
		
		%ellipsis 
		% from user3 to userN 
		\path (4,0.1875) -- node[auto=false]{\ldots} (5,0.1875);
		\path (1.15, 1.675) -- node[auto=false]{\ldots} (2.15, 1.1675);
	\end{tikzpicture}
	\caption{An illustration of uplink NOMA with $N$ users.
		\vspace{-0.6cm}}
	\label{Fig:NOMA}
\end{figure}
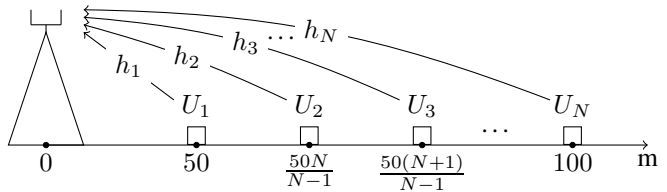

The system model considered here represents an uplink NOMA-based network scenario (depicted in Fig. \ref{Fig:NOMA}) comprising BS and $N$ end-users, denoted by $U_{k}$, with $k\in\mathcal{A} = \{1,2,\ldots,N\}$. It is assumed that all nodes are deployed with a single antenna. The channel between the user of interest and BS is given by $\tilde{h}_k$ and assumed to be independent and identically distributed random variables (RVs) following Rayleigh fading.

\subsection{Non-orthogonal Multiple Access}
Due to the considered uplink mode of communication, all end-users simultaneously transmit their individual messages to BS while ensuring the NOMA principle to be true through the path loss associated with their corresponding distances\footnote{Note that the similar assumption can be made using the principle of messages' priorities.}. Hence, the received signal at BS can be written as
\begin{align}
	\label{y}
	y = \sum_{k=1}^{N} s_{k} h_{k} + n,
\end{align}
where $s_{k}$ is the transmitted message of user $k$, with the transmit power defined as $\mathbb{E}\{s_{k} s_{k}^{*}\} = P_k$. For the sake of simplicity, we assume that all end-users transmit with the same power levels, {\it i.e.}, $P = P_1 = P_2 = \ldots = P_N$. $h_{k}$ is modelled as $h_{k} = \tilde{h}_k / \sqrt{d_k^{\tau}}$, where $d_k$ and $\tau$ indicate the corresponding distance and path-loss exponent. $n$ is an additive white Gaussian noise (AWGN) term, with zero mean and variance $\sigma_n^2$. 

\subsection{Maximum Likelihood Detection}
The receiver implements the ML detection as a decoding method, which output, given by an optimal symbol vector $\hat{\mathbf{s}}^{}_{ML}$, is formulated as 
\begin{equation}
	\label{ml}
	\hat{\mathbf{s}}^{}_{ML} = \argmin_{\hat{s}_1, \hat{s}_2,\dots \hat{s}_N}|y - \mathbf{s}\mathbf{h}^{\mathrm{T}}|^{2},
\end{equation}
where $y$ is the received signal, $\mathbf{s}$ and $\mathbf{h}$ are the $ 1\times N $ vectors representing the transmit symbols and channel gains of end-users, respectively. 

\subsection{Quantum Annealing}
Quantum computing implements the concepts of quantum mechanics to solve high-computational problems faster than classical computers. There are several types of quantum computers. If the quantum circuit model controls the qubits using quantum logic gates to solve a wide range of problems, QA is an effective tool to solve optimization problems with given objective functions. Specifically, the quantum processing unit (QPU) uses quantum fluctuations to search for the global minimum solution. However, modern quantum computers are not that perfect yet and QA might end up with a local minimum solution due to the analogue noises in the device \cite{dwave_doc_errors}. To mitigate the internal errors in QPU, the annealing parameters could be modified based on the specifics of the problem. To minimize these errors, QA can be simulated many times, and the optimal solution will be the answer with the lowest energy level and most frequent occurrences. The annealing parameter that controls the number of samples per simulation is the \textit{number of reads} ($R$), it was decided to set it to $1000$. Another annealing parameter that could impact the performance of the QA machine is the \textit{single-sample annealing time} ($T_a$), which was set to $20$  $\mu s$  without any pausing time \cite{dwave_doc_parameters}. 

%Today D-Wave offers D-Wave 2000Q and D-Wave Advantage systems, the former is an old version with 2048 qubits on a Chimera graph architecture, while the latter is a recently released model that proposes 5640 available physical qubits on a Pegasus topology. The quantum annealers can be accessed through Amazon Web Services or D-Wave's Leap cloud platform. 

\subsection{Quadratic Unconstrained Binary Optimization}
To solve an optimization problem by a means of QA, the objective function should be reformulated in a form of the quadratic unconstrained binary optimization (QUBO) model or the Ising model. These formulations can be used interchangeably, and it is easy to switch from one to another. At the end of annealing process, the best solution is given by the qubit sequence with a minimum energy state. The model is given by an upper-diagonal matrix $Q$ with a size of $ M\times M $
\begin{equation}
	\label{qubo}
	\hat{q}_1,\hat{q}_2,\ldots,\hat{q}_M = \argmin _{q_1, q_2,\dots q_M}\sum_{i\le j}^{M}Q_{ij}q_{i}q_{j},
\end{equation}
where $M$ denotes the number of qubits, and each variable $q_{i}$ takes binary numbers. By applying the property $q_{i}^{2} = q_{i}q_{i} = q_{i}$, and substituting the symbols in \eqref{ml} by the corresponding binary equivalent form, it becomes feasible to formulate the NOMA ML problem in terms of the QUBO form. The corresponding diagonal $Q_{ii}$ and non-diagonal $Q_{ij}$ terms of the QUBO model are represented by linear and quadratic coefficients, accordingly.

To implement the obtained QUBO model on QPU, the matrix coefficients have to be mapped into physical qubits of a QA chip. The linear coefficients define the qubit biases, while quadratic coefficients represent the coupling strengths \cite{dwave_doc_embed}. However, QPU's physical qubits are not fully connected, in the Chimera graph each qubit has six neighbours, and, in Pegasus topology, each qubit is connected with $15$ neighbour qubits \cite{dwave_doc_topologies}. Therefore, sometimes it is not possible to directly apply the QUBO model onto real QA machine. 
%Embedding is a process that integrates logic binary variables from \eqref{qubo} into physical qubits, by describing one logical variable with several physical qubits.  

\begin{figure}[!t]
	\centering
	\includegraphics[width=0.9\columnwidth]{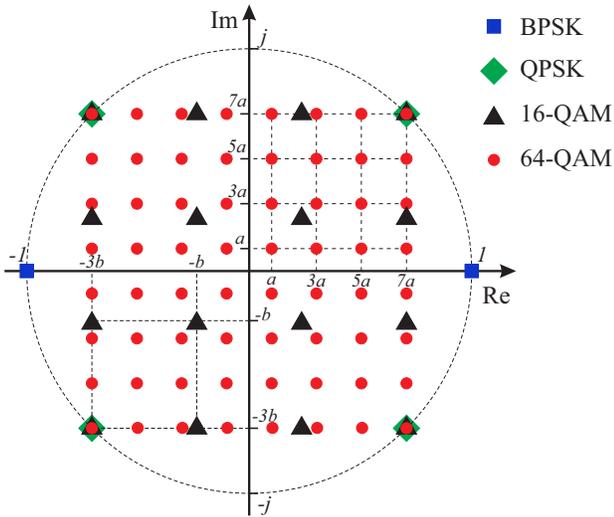}
	\caption{Constellations of different modulation types.
	\vspace{-0.5cm}}
	\label{fig:modulation}
\end{figure}

\section{Formulation of NOMA ML Problem in a QUBO model}
\label{section:formulation}
The ML expression in \eqref{ml} should be described by a QUBO model to solve it with QA. Taking into account \eqref{y} and the complex numbers due to the broadcast transmission, we split the received signal $y$ and channel coefficients into real and imaginary parts as follows: $y = y_{\rm R} + j y_{\rm I}$ and $h_{k} = h_{k,{\rm R}} + j h_{k,{\rm I}}$. Each symbol, denoted by $s_{k}$, can be transformed into the qubit form. The methodology for representing symbols in terms of the binary variables depends on the modulation type. The considered modulations are presented in Fig. \ref{fig:modulation}.

\subsubsection{BPSK}
The BPSK symbols of a form $ s_{k}\in\{\pm 1\} $, can be reformulated into the QUBO form by $s_{k} = 2q_{k} - 1$. In this case, $M$ in \eqref{qubo} takes a value of the number of users, $N$.

\subsubsection{QPSK} As shown in Fig. \ref{fig:modulation}, the QPSK symbols, $ s_{k} \in \{\frac{\pm1 \pm j}{\sqrt{2}}\} $, reside on a unit circle with $90^{\circ}$ phase difference between two consecutive symbols. They can be represented by two qubits; therefore, the term $M$ in \eqref{qubo} becomes the double of a number of users, $2N$. Finally, the QPSK symbols can be expressed as $ s_{k} = \left[\left(2q_{2k-1}-1\right) + j\left(2q_{2k}-1\right)\right]/\sqrt{2}$.

\subsubsection{16-QAM} 
The $16$-QAM modulation encodes four information bits into a complex symbol and correspondingly includes $16$ constellation points. The symbols $ s_{k}\in\{\pm b\pm jb, \pm b\pm j3b, \pm 3b\pm jb, \pm 3b \pm j3b\} $, where $b = 1/(3\sqrt{2})$, require four qubits to be adapted for the QUBO model, i.e., $ s_{k} = \left[\left(4q_{4k-3}+2q_{4k-2}-3\right) + j\left(4q_{4k-1}+2q_{4k}-3\right)\right]/3\sqrt{2}$.
$M$ in \eqref{qubo} becomes the quadruple of a number of users, $4N$.

%For greater efficiency, the Gray coding can be used as an encoding technique at the transmitter side, and the QUBO solution bits should then be transformed from QuAMax into the Gray code mapping form. 

\subsubsection{64-QAM}
On the other hand, the $64$-QAM modulation returns a complex symbol by encoding $16$ information bits. As shown in Fig. \ref{fig:modulation}, the constellation points are defined as per Fig. \ref{fig:modulation} 
%\begin{equation*}
%	s_{k} \hspace{-0.1cm} \in \hspace{-0.1cm} \left\lbrace 
%	\begin{array}{cccc}
%		\pm a\pm ja & \pm a\pm j3a & \pm a\pm j5a & \pm a\pm j7a \\
%		\pm 3a\pm ja & \pm 3a \pm j3a & \pm 3a\pm j5a & \pm 3a \pm j7a\\
%		\pm 5a\pm ja & \pm 5a \pm j3a & \pm 5a\pm j5a & \pm 5a \pm j7a \\
%		\pm 7a\pm ja & \pm 7a \pm j3a & \pm 7a\pm j5a & \pm 7a \pm j7a
%	\end{array} \right\rbrace,
%\end{equation*}
%where $a = 1/(7\sqrt{2})$, 
and require six qubits to be adopted for the QUBO model; therefore, the term $M$ in \eqref{qubo} becomes the sixtuple of a number of users, $6N$.
These symbols can be re-expressed as $ s_{k} = a \left(8q_{6k-5}+4q_{6k-4}+2q_{6k-3}-7\right) + j a \left(8q_{6k-2}+4q_{6k-1}+2q_{6k}-7\right)$.

%\begin{figure}[!t]
%	\centering
%	\includegraphics[width=\columnwidth]{results1}
%	\caption{The BER performance vs. the transmit SNR for a two-user NOMA scenario, with different modulation types.}
%\end{figure}

\begin{figure}[!t]
	\centering
	\includegraphics[width=0.9\columnwidth]{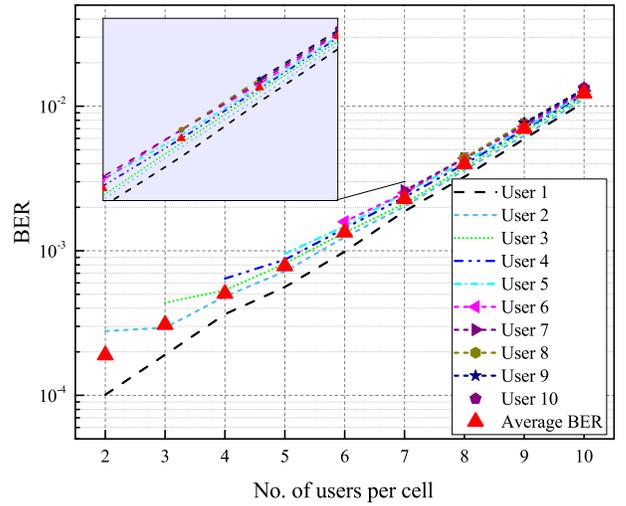}
	\caption{BER vs. the number of users per cell under BPSK.
	\vspace{-0.5cm}}
	\label{fig:multi-user NOMA}
\end{figure}

\begin{figure}[!t]
	\centering
	\includegraphics[width=1\columnwidth]{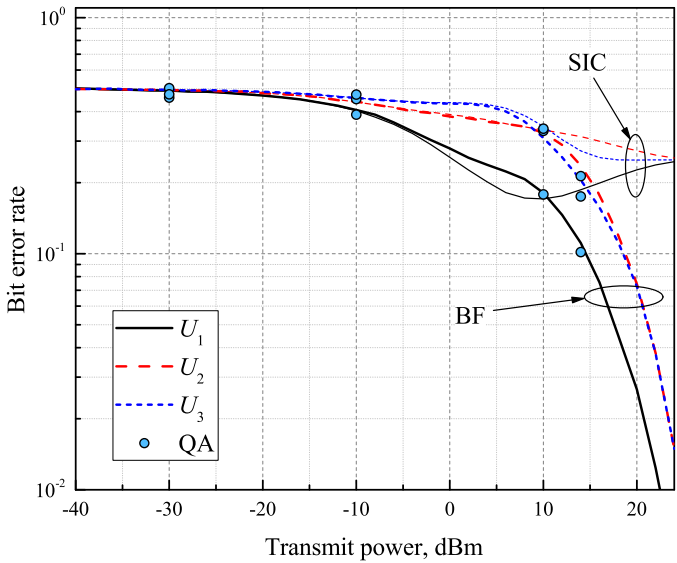}
	\caption{The BER performance vs. the transmit power for a three-user NOMA scenario, with different decoding techniques.
	\vspace{-0.5cm}}
	\label{fig:3-user}
\end{figure}

\begin{figure}[!t]
	\centering
	\includegraphics[width=1\columnwidth]{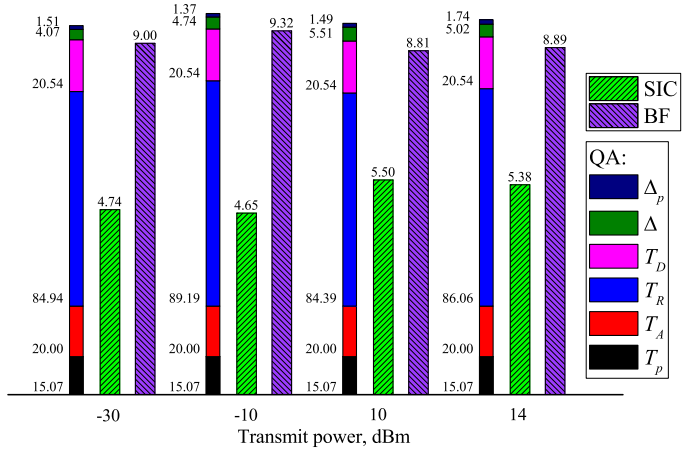}
	\caption{Comparison between the decoding techniques in terms of simulation time for five ($5$) samples (in $ms$).
	\vspace{-0.5cm}}
	\label{fig:Timing}
\end{figure}

\section{Implementation and Results Discussion}
\label{section:implementation}
\subsection{Implementation}
The system model considered in the analysis represents a $N$-end user uplink NOMA system that employs BPSK modulation. As illustrated in Fig. \ref{Fig:NOMA}, the base station is located at a coordinate $0$. We assume that the first and the farthest users are located $50$ and $100$ m away from the base station, respectively, while the other intermediate nodes are evenly distributed in-between. For practical purposes, we apply the standardized scenarios described in \cite{tech2010} and \cite{tech2017}. The performance of the exploited solvers (i.e., BF, SIC and QA) were examined by varying the transmit power from $-40$ dBm to $24$ dBm. Moreover, the environment of data transmission is considered to be a free space with line-of-sight propagation which relates to the path-loss exponent $\tau = 2$ \cite{pathloss}.

The QA method was implemented on the proposed system model using D-Wave's Advantage QPU \cite{dwave_doc_topologies}. It is important to mention that the parallelization procedure was used to save the simulation time on the QA computer \cite{leveraging}, i.e., $5$ instances of the problem were run at the same time on QPU. Since the $3$-user BPSK problem requires only $3$ logical qubits, $5$ instances would occupy $15$ logical qubits. In the analysis of simulation time, the following terms are used as follows. \textit{QPU Programming Time} ($T_p$) is the time taken for programming the couplers and biases of the chip in accordance with the QUBO model. \textit{QPU Sampling Time} ($T_s$) is the total time for simulation of $R$ samples and consists of the annealing time ($T_a$), the readout time ($T_r$), and the delay time ($T_d$) for every single sample \cite{qpu_timing}. For the sake of simplicity, we declare the following terms: the \textit{total QPU Annealing Time} ($T_A = R\cdot T_a$), the \textit{total QPU Readout Time} ($T_R = R\cdot T_r$), and the \textit{QPU Delay Time} ($T_D = R\cdot T_d$). $T_p$, $T_s$, \textit{QPU Access Overhead Time} ($\Delta$) and \textit{Post-processing Overhead Time} ($\Delta_p$) constitute \textit{QPU Service Time}, which is considered as the total time taken for the QA simulation excluding internet delay.

%\colorbox{blue}{The BF and SIC methods were executed on a PC with $ 2.6 $ GHz Intel Core i$ 7 $ processor and $8$ GB random access memory.}

%The \textit{QPU Service Time} is considered as the total time taken for QA simulation excluding internet delay. It can be broken down into the \textit{QPU Access Time}, \textit{QPU Access Overhead Time} ($\Delta$) and \textit{Post-processing Overhead Time} ($\Delta_p$). The \textit{QPU Access Time} itself can be decomposed into \textit{QPU Programming Time} ($T_p$) and \textit{QPU Sampling Time} ($T_s$). $T_p$ is the time taken for programming the couplers and biases of the chip in accordance with the QUBO model. $T_s$ is the total time for simulation of $R$ samples, it consists of the annealing time ($T_a$), the readout time ($T_r$) and the delay time ($T_d$) for every single sample \cite{qpu_timing}. For the sake of simplicity, we declare the following terms: the \textit{total QPU Annealing Time} ($T_A = R\cdot T_a$), the \textit{total QPU Readout Time} ($T_R = R\cdot T_r$) and the \textit{QPU Delay Time} ($T_D = R\cdot T_d$). 

\begin{table}[t!]
	\begin{center}
		\footnotesize
		\caption{Simulation time (in $ms$) of the decoding techniques.}
		\label{tab:table1}
		\begin{tabular}{l|c|r}
			\textbf{Technique} & \textbf{for $5$ samples} & \textbf{for $1$ sample}\\
			\hline
			SIC &  $5.066$ &  $3.958$\\
			\hline
			BF & $9.006$ & $4.535$\\ 
			\hline
			QA &   & \\ 
			QPU Service Time &  $148.116$ & $107.984$\\ 
			\hdashline
			\quad $T_p$ & $15.069$ & $15.068$\\
			\quad $T_A$ & $20$ & $20$\\
			\quad $T_R$ & $86.145$ & $46.68$\\
			\quad $T_D$ & $20.54$ & $20.54$\\
			\quad $\Delta$ & $4.835$ & $3.86$\\
			\quad $\Delta_p$ & $1.527$ & $1.836$\\
		\end{tabular}
	\end{center}
\vspace{-0.6cm}
\end{table}

\subsection{Discussion}

In Fig. \ref{fig:multi-user NOMA}, we examine the influence of the number of users per cell on the BER performance in the NOMA system. Particularly, the transmit power and the AWGN power were set to $10$ dBm and $-60$ dBm, respectively. As expected, the average BER performance degrades with the increase in the number of users per cell. It is seen from the plot that, for the number of users greater than $6$, the difference between the BER curves becomes quite similar to each other. Therefore, we apply the $3$-user NOMA scenario in the further analysis.

In Figs. \ref{fig:3-user} and \ref{fig:Timing}, we compare the SIC, BF and QA decoding techniques in terms of the BER performance and simulation time, with the noise power set to $\sigma_n^2 = -30$ dBm. QA was simulated for particular transmit power levels, i.e., $P =\{-30, -10, 10, 14\}$ dBm. The number of QA simulations per power level was taken to ensure a $1\%$ accuracy with respect to the BER performance of BF. Overall, about $380$ simulations (with $5$ problems at a time) were performed on QPU. 

%Fig. 4
Fig. \ref{fig:3-user} illustrates the BER performances of each decoding method. The BER curves can be characterized by different behaviour. It can be seen from the $U_1$'s curve, the SIC and BF performances coincide up to $-5$ dBm, and then SIC starts outperforming BF over a short transmit power range (i.e., from $-5$ dBm to $10$ dBm). However, afterwards, SIC starts experiencing noticeable performance degradation. At the same time, $U_2$ has identical BER for SIC and BF up to $10$ dBm. After this point, there is a slight improvement of the SIC curve, but, in general, its performance is much worse than that of BF. What stands out for $U_3$ is the equal performance of SIC and BF for transmit power less than $5$ dBm, thereafter, the difference between them begins to grow considerably. The performance of BF improves remarkably after $5$ dBm, whereas the SIC curve enhances minimally until $15$ dBm with the following saturation after this point. In general, the QA, BF and SIC curves follow the same trend till $10$ dBm, but after this level, the SIC performance degrades substantially, while the BER result of QA is approximately the same as BF. The reason for the SIC's low performance could be insufficient differences between the end-users' power levels.

%Fig. 5
Fig. \ref{fig:Timing} presents the timing data for each technique. As can be seen from the histogram plot, SIC shows the fastest execution among the presented techniques. As expected, the second fastest technique is BF, since it needs to iterate over all possible combinations. The QA execution time interval consists of several timing sub-intervals. While $T_p$, $T_A$ and $T_D$ are the same for all cases, $T_R$ shows a slight variation, since the reading time depends on the position of exploited physical qubits on the chip and on the number of used qubits (more time is needed for a larger number of qubits) \cite{qpu_timing}. The dependence on the number of qubits could be noticed in Table \ref{tab:table1}, i.e., $T_R$ for $1$ sample is twice as fast as $T_R$ for $5$ samples. Overall, the total time taken for QA execution is about $30$ times more than the SIC simulation. However, it is important to mention that QA needs to simulate the same problem $R$ number of times (in our case it is set to $1000$). Furthermore, the problem could be addressed by the parallelization method. From Table \ref{tab:table1}, it is clearly noticeable that the execution of $5$ instances in parallel is more than $3$ times faster than executing the same number of samples in a sequential order. Therefore, one could potentially conclude that the overall QA simulation time can be decreased by simulating multiple problem instances at the same time. 

%According to Figs. \ref{fig:3-user} and \ref{fig:Timing}, the SIC results in the minimum execution time, but the BER performance of the method is the worst among the presented methods.  While the BF took more time than SIC to execute, as expected, since BF needs to iterate over all possible combinations. However, the BER performance of the technique is the most optimal one. One can observe that the BER result of the QA is close to the BF performance. But the QA simulation is the slowest considering that it needs time for programming QPU, executing the same problem multiple times, and other procedures needed to mitigate internal errors. 

% It should be noted that the actual duration of the computer's computation is $T_A$ \cite{leveraging}. Hopefully, with the new advancements in quantum computation, the impact of internal noises on the solution will be reduced, and the need of performing error mitigating procedures will be vanished. This will pave a way for the execution only single-sample annealing $T_a$, which takes up several $\mu s$.

\section{Conclusion}
\label{section:conclusion}
To mitigate the problem of spectrum scarcity and present an alternative for SIC, this work aims at evaluating the performance of QA-aided ML detection for NOMA systems. The ML problem was firstly described in terms of the QUBO model to enable the integration of the problem with QPU. For $3$-user NOMA under BPSK modulation, the BER performance of QA is approximately the same as the BF method, but QA takes longer time to execute due to the current hardware specifics. The parallelization method seems to be a potential solution that could decrease the execution time of QA. Additionally, the analogous noises in QA could be suppressed with coming of the next-generation QA hardware, and the time taken for error alleviation would be reduced. This will pave the way for computing time-dependent operations on QA including the QA-assisted ML decoding technique for future NOMA systems. %preparation for QA computing

\section{Acknowledgement}
This work was supported by the Nazarbayev University (NU) Social Policy grant, the NU FDCRP Grant no. 240919FD3935 and the NU CRP Grant no. 11022021CRP1513.

\appendices
\numberwithin{equation}{section}
\section{QUBO Model Coefficients} 
In this section, we present the coefficients for the QUBO model for different modulation types\footnote{Note that the equations involving the positive integer variables $i$ and $j$ must satisfy the condition $i<j$.}.
\subsubsection{BPSK}
The corresponding $Q_{ij}^{\rm B}$ values can be found as
\begin{align}
	\label{bpsk_Q_ii}
	Q_{ii}^{\rm B} &= P \left[ -4 h_{i,{\rm R}} \left( \sum_{l=1}^{N} h_{l,{\rm R}} \hspace{-0.1cm} - \hspace{-0.1cm} h_{i,{\rm R}} \right) 
	\hspace{-0.1cm} - \hspace{-0.1cm} 4 h_{i,{\rm I}} 
	\left( \sum_{l=1}^{N} h_{l,{\rm I}} \hspace{-0.1cm} - \hspace{-0.1cm} h_{i,{\rm I}} \right)
	\right] \nonumber\\
	&~~~ - \sqrt{P} \left( 4 y_{\rm R} h_{i,{\rm R}} 
	+ 4 y_{\rm I} h_{i,{\rm I}} \right), \\
	\label{bpsk_Q_ij}
	Q_{ij}^{\rm B} &= P \left( 8 h_{i,{\rm R}} h_{j,{\rm R}} 
	+ 8 h_{i,{\rm I}} h_{j,{\rm I}} \right).
\end{align}
\subsubsection{QPSK}
The $Q_{ij}^{\rm Q}$ values can be found using \eqref{qpsk_Q_ii_odd}, \eqref{qpsk_Q_ii_even} and the equations given below
\begin{figure*}[!t]
	\small
	\begin{align}
		\label{qpsk_Q_ii_odd}
		Q_{(2i-1),(2i-1)}^{\rm Q} &= \frac{P}{2} \left[ -4 h_{i,{\rm R}} \left(
		\left( \sum_{l=1}^{N} h_{l,{\rm R}} - h_{i,{\rm R}} \right) - \left( \sum_{l=1}^{N} h_{l,{\rm I}} - h_{i,{\rm I}} \right)
		\right) - 4 h_{i,{\rm I}} 
		\left(
		\left( \sum_{l=1}^{N} h_{l,{\rm R}} - h_{i,{\rm R}} \right) + 
		\left( \sum_{l=1}^{N} h_{l,{\rm I}} - h_{i,{\rm I}} \right)
		\right)
		\right] \nonumber\\
		&~~~ - \sqrt{2 P} \left( 2 y_{\rm R} h_{i,{\rm R}} 
		+ 2 y_{\rm I} h_{i,{\rm I}} \right)
	\end{align}
	\begin{align}
		\label{qpsk_Q_ii_even}
		Q_{(2i),(2i)}^{\rm Q} &= \frac{P}{2} \left[ -4 h_{i,{\rm R}} \left(
		\left( \sum_{l=1}^{N} h_{l,{\rm R}} - h_{i,{\rm R}} \right) + \left( \sum_{l=1}^{N} h_{l,{\rm I}} - h_{i,{\rm I}} \right)
		\right) - 4 h_{i,{\rm I}} 
		\left(-
		\left( \sum_{l=1}^{N} h_{l,{\rm R}} - h_{i,{\rm R}} \right) + 
		\left( \sum_{l=1}^{N} h_{l,{\rm I}} - h_{i,{\rm I}} \right)
		\right)
		\right] \nonumber\\
		&~~~ - \sqrt{2P} \left( -2 y_{\rm R} h_{i,{\rm I}} 
		+ 2 y_{\rm I} h_{i,{\rm R}} \right) 
	\end{align}
	\hrule
\end{figure*}
{\allowdisplaybreaks
\begin{align}
	\label{qpsk_Q_odd_i_even_i}
	Q_{(2i-1),(2i)}^{\rm Q} &= 0,\quad \forall i\ge1, \\
	\label{qpsk_Q_odd_ij_even_ij}
	Q_{(2i-1),(2j-1)}^{\rm Q} &= Q_{(2i),(2j)}^{\rm Q} = \frac{P}{2} \left( 8 h_{i,{\rm R}} h_{j,{\rm R}} + 8 h_{i,{\rm I}} h_{j,{\rm I}} \right), \\
	\label{qpsk_Q_odd_i_even_j}
	Q_{(2i-1),(2j)}^{\rm Q} &= \frac{P}{2} \left( 8 h_{i,{\rm I}} h_{j,{\rm R}} 
	- 8 h_{i,{\rm R}} h_{j,{\rm I}} \right), \\
	\label{qpsk_Q_even_i_odd_j}
	Q_{(2i),(2j-1)}^{\rm Q} &= \frac{P}{2} \left( -8 h_{i,{\rm I}} h_{j,{\rm R}} 
	+ 8 h_{i,{\rm R}} h_{j,{\rm I}} \right).
\end{align}}
\subsubsection{16-QAM}
The $Q_{ij}^{16{\rm Q}}$ values can be found using
\begin{figure*}[!t]
	\vspace{-0.5cm}
	\small
	\begin{multline}
		\label{16q_Q_4i_minus_3_both}
		Q_{(4i-3),(4i-3)}^{16{\rm Q}} = \frac{P}{9} \left[ 
		-4 h_{i,{\rm R}} \left( h_{i,{\rm R}} + 3 \left(  \left( \sum_{k=1}^{N} h_{k,{\rm R}} - h_{i,{\rm R}} \right) - \left( \sum_{k=1}^{N} h_{k,{\rm I}} - h_{i,{\rm I}} \right)
		\right) \right)
		\right. \\
		\left.
		-4 h_{i,{\rm I}} \left( h_{i,{\rm I}} + 3 \left(  \left( \sum_{k=1}^{N} h_{k,{\rm R}} - h_{i,{\rm R}} \right) + \left( \sum_{k=1}^{N} h_{k,{\rm I}} - h_{i,{\rm I}} \right)
		\right) \right) 
		\right] 
		- \frac{\sqrt{2 P}}{3}\left( 4 y_{\rm R} h_{i,{\rm R}} + 4 y_{\rm I} h_{i,{\rm I}} \right)
	\end{multline}
	\begin{multline}
		\label{16q_Q_4i_minus_2_both}
		Q_{(4i-2),(4i-2)}^{16{\rm Q}} = \frac{P}{9} \left[ 
		-4 h_{i,{\rm R}} \left( h_{i,{\rm R}} + \frac{3}{2} \left(  \left( \sum_{k=1}^{N} h_{k,{\rm R}} - h_{i,{\rm R}} \right) - \left( \sum_{k=1}^{N} h_{k,{\rm I}} - h_{i,{\rm I}} \right)
		\right) \right)
		\right. \\
		\left.
		-4 h_{i,{\rm I}} \left( h_{i,{\rm I}} + \frac{3}{2} \left(  \left( \sum_{k=1}^{N} h_{k,{\rm R}} - h_{i,{\rm R}} \right) + \left( \sum_{k=1}^{N} h_{k,{\rm I}} - h_{i,{\rm I}} \right)
		\right) \right) 
		\right] 
		- \frac{\sqrt{2 P}}{3}\left( 2 y_{\rm R} h_{i,{\rm R}} + 2 y_{\rm I} h_{i,{\rm I}} \right)
	\end{multline}
	\begin{multline}
		\label{16q_Q_4i_minus_1_both}
		Q_{(4i-1),(4i-1)}^{16{\rm Q}} = \frac{P}{9} \left[ 
		-4 h_{i,{\rm R}} \left( h_{i,{\rm R}} + 3 \left(  \left( \sum_{k=1}^{N} h_{k,{\rm R}} - h_{i,{\rm R}} \right) + \left( \sum_{k=1}^{N} h_{k,{\rm I}} - h_{i,{\rm I}} \right)
		\right) \right)
		\right. \\
		\left.
		-4 h_{i,{\rm I}} \left( h_{i,{\rm I}} + 3 \left( - \left( \sum_{k=1}^{N} h_{k,{\rm R}} - h_{i,{\rm R}} \right) + \left( \sum_{k=1}^{N} h_{k,{\rm I}} - h_{i,{\rm I}} \right)
		\right) \right) 
		\right] 
		- \frac{\sqrt{2 P}}{3}\left( -4 y_{\rm R} h_{i,{\rm I}} + 4 y_{\rm I} h_{i,{\rm R}} \right)
	\end{multline}
	\begin{multline}
		\label{16q_Q_4i_both}
		Q_{(4i),(4i)}^{16{\rm Q}} = \frac{P}{9} \left[ 
		-4 h_{i,{\rm R}} \left( h_{i,{\rm R}} + \frac{3}{2} \left(  \left( \sum_{k=1}^{N} h_{k,{\rm R}} - h_{i,{\rm R}} \right) + \left( \sum_{k=1}^{N} h_{k,{\rm I}} - h_{i,{\rm I}} \right)
		\right) \right)
		\right. \\
		\left.
		-4 h_{i,{\rm I}} \left( h_{i,{\rm I}} + \frac{3}{2} \left( - \left( \sum_{k=1}^{N} h_{k,{\rm R}} - h_{i,{\rm R}} \right) + \left( \sum_{k=1}^{N} h_{k,{\rm I}} - h_{i,{\rm I}} \right)
		\right) \right) 
		\right] 
		- \frac{\sqrt{2 P}}{3}\left( -2 y_{\rm R} h_{i,{\rm I}} + 2 y_{\rm I} h_{i,{\rm R}} \right)
	\end{multline}
	\hrule
\end{figure*}
{\allowdisplaybreaks
	\begin{align}
	\label{16q_Q_zero}
	Q_{(4i-3),(4i-1)}^{16{\rm Q}} &= Q_{(4i-3),(4i)}^{16{\rm Q}}  \nonumber \\
	&= Q_{(4i-2),(4i-1)}^{16{\rm Q}} = Q_{(4i-2),(4i)}^{16{\rm Q}} = 0, \\
	\label{16q_Q_odd_even}
	Q_{(4i-3),(4i-2)}^{16{\rm Q}} &= Q_{(4i-1),(4i)}^{16{\rm Q}} = \frac{8 P}{9} \left( |h_{i,{\rm R}}|^2 + |h_{i,{\rm I}}|^2 \right), \\
	\label{16q_Qij_Eq7}
	Q_{(4i-3),(4j-2)}^{16{\rm Q}} &= Q_{(4i-2),(4j-3)}^{16{\rm Q}} = Q_{(4i),(4j-1)}^{16{\rm Q}} \nonumber \\
	= Q_{(4i-1),(4j)}^{16{\rm Q}} &= 
	\frac{P}{9} \left( 8 h_{i,{\rm R}} h_{j,{\rm R}} + 8 h_{i,{\rm I}} h_{j,{\rm I}} \right), \\
	\label{16q_Qij_even}
	Q_{(4i-2),(4j-2)}^{16{\rm Q}} &= Q_{(4i),(4j)}^{16{\rm Q}} = \frac{P}{9} \left( 4 h_{i,{\rm R}} h_{j,{\rm R}} + 4 h_{i,{\rm I}} h_{j,{\rm I}} \right), \\
	\label{16q_Qij_odd}
	Q_{(4i-1),(4j-1)}^{16{\rm Q}} &= Q_{(4i-3),(4j-3)}^{16{\rm Q}}  \nonumber\\
	&= \frac{P}{9} \left( 16 h_{i,{\rm R}} h_{j,{\rm R}} + 16 h_{i,{\rm I}} h_{j,{\rm I}} \right), \\
	\label{16q_Qij_Eq8}
	Q_{(4i-2),(4j-1)}^{16{\rm Q}} &= Q_{(4i-3),(4j)}^{16{\rm Q}} = \frac{8 P}{9} \left( - h_{i,{\rm R}} h_{j,{\rm I}} + h_{i,{\rm I}} h_{j,{\rm R}} \right), \\
	\label{16q_Qij_Eq9}
	Q_{(4i-1),(4j-2)}^{16{\rm Q}} &= Q_{(4i),(4j-3)}^{16{\rm Q}} \nonumber \\
	&= \frac{P}{9} \left( -8 h_{i,{\rm I}} h_{j,{\rm R}} + 8 h_{i,{\rm R}} h_{j,{\rm I}} \right), \\
	\label{16q_Qij_Eq10}
	Q_{(4i-2),(4j)}^{16{\rm Q}} &= 
	\frac{P}{9} \left( -4 h_{i,{\rm R}} h_{j,{\rm I}} + 4 h_{i,{\rm I}} h_{j,{\rm R}} \right), \\
	\label{16q_Qij_Eq11}
	Q_{(4i),(4j-2)}^{16{\rm Q}} &= 
	\frac{P}{9} \left( -4 h_{i,{\rm I}} h_{j,{\rm R}} + 4 h_{i,{\rm R}} h_{j,{\rm I}} \right), \\
	\label{16q_Qij_Eq12}
	Q_{(4i-1),(4j-3)}^{16{\rm Q}} &= 
	\frac{P}{9} \left( -16 h_{i,{\rm I}} h_{j,{\rm R}} + 16 h_{i,{\rm R}} h_{j,{\rm I}} \right), \\
	\label{16q_Qij_Eq13}
	Q_{(4i-3),(4j-1)}^{16{\rm Q}} &= 
	\frac{P}{9} \left( -16 h_{i,{\rm R}} h_{j,{\rm I}} + 16 h_{i,{\rm I}} h_{j,{\rm R}} \right).
\end{align}}

\subsubsection{64-QAM}
The derivations of $Q_{ij}^{64{\rm Q}}$ values are omitted due to the submission page limit and will be included in the extended version, if accepted.

\balance


\begin{thebibliography}{00}
\bibitem{cisco} 
``Cisco Annual Internet Report - Cisco Annual Internet Report (2018–2023) White Paper," Cisco, March 2020, [Online]. Available: https://www.cisco.com/c/en/us/solutions/collateral/executive-perspectives/annual-internet-report/white-paper-c11-741490.html.
	
\bibitem{ldpc}
S. Kasi and K. Jamieson, ``Towards quantum belief propagation for LDPC decoding in wireless networks," {\it 26th Annu. Int. Conf. Mobile Comput. Netw.,} London, UK, pp. 1-14, Sep. 2020.

%\bibitem{vpp} 
%S. Kasi, A. K. Singh, D. Venturelli, and K. Jamieson, ``Quantum Annealing for Large MIMO Downlink Vector Perturbation Precoding," {\it IEEE International Conference on Communications,} Montreal, Canada, pp. 1-6, June 2021. 

\bibitem{leveraging} 
M. Kim, D. Venturelli, and K. Jamieson, ``Leveraging Quantum Annealing for Large MIMO Processing in Centralized Radio Access Networks," in {\it Proc. of the ACM Special Interest Group on Data Communication,} pp. 241-255, Aug. 2019.

\bibitem{tabi} 
Z. I. Tabi {\it et al.}, 
%Z. I. Tabi, Á. Marosits, Z. Kallus, P. Vaderna, I. Gódor, and Z. Zimborás, 
``Evaluation of Quantum Annealer Performance via the Massive MIMO Problem," {\it IEEE Access,} vol. 9, pp. 131658-131671, 2021.

\bibitem{kizilirmak} 
R. C. Kizilirmak, ``Quantum Annealing Approach to NOMA Signal Detection," {\it 12th Int. Symp. Commun. Syst. Netw. Digital Signal Process. (CSNDSP),} Porto, Portugal, pp. 1-5, Jul. 2020.

\bibitem{tdma} 
F. Ishizaki, ``Computational Method Using Quantum Annealing for TDMA Scheduling Problem in Wireless Sensor Networks," {\it 13th Int. Conf. Signal Process. Commun. Syst. (ICSPCS),} Australia, pp. 1-9, Dec. 2019.

\bibitem{noma_tutorial}  
R. C. Kizilirmak, ``Non-Orthogonal Multiple Access (NOMA) for 5G Networks", {\it Towards 5G Wireless Networks - A Physical Layer Perspective}. London, United Kingdom: IntechOpen, 2016 [Online]. Available: https://www.intechopen.com/chapters/52822 

\bibitem{survey_noma}   
B. Makki {\it et al.}, 
% B. Makki, K. Chitti, A. Behravan and M.-S. Alouini,
``A Survey of NOMA: Current Status and Open Research Challenges," {\it IEEE Open J. Commun. Soc.,} vol. 1, pp. 179-189, 2020.

\bibitem{dwave_doc_errors} 
``Error Sources for Problem Representation" {\it D-Wave System Documentation}, 2021 [Online]. Available: https://docs.dwavesys.com/docs/latest/c\_qpu\_ice.html [Accessed Apr. 2022].

\bibitem{dwave_doc_parameters} 
``Solver Parameters," {\it D-Wave System Documentation}, 2021 [Online]. Available: https://docs.dwavesys.com/docs/latest/c\_solver\_parameters.html [Accessed Apr. 2022].

\bibitem{dwave_doc_embed} 
``Constraints Example: Minor-Embedding" {\it D-Wave System Documentation}, 2021 [Online]. Available: https://docs.dwavesys.com/docs/latest/c\_gs\_7.html[Accessed Apr. 2022].

\bibitem{dwave_doc_topologies} 
``D-Wave QPU Architecture: Topologies" {\it D-Wave System Documentation}, 2021 [Online]. Available: https://docs.dwavesys.com/docs/latest/c\_gs\_7.html[Accessed Apr. 2022].

\bibitem{tech2010} 
``LTE; Evolved Universal Terrestrial Radio Access (E-UTRA); Base Station (BS) radio transmission and reception (3GPP TS 36.104 version 9.4.0 Release 9)," ETSI, France, Tech. Specification. RTS/TSGR-0436104v940, Jul. 2010. 

\bibitem{tech2017} 
``LTE; Evolved Universal Terrestrial Radio Access (E-UTRA); User Equipment (UE) radio transmission and reception (3GPP TS 36.101 version 14.3.0 Release 14)," ETSI, France, Tech. Specification. RTS/TSGR-0436101ve30, Apr. 2017. 

\bibitem{pathloss} 
T. Rappaport, ``Mobile Radio Propagation: Large-Scale Path Loss," \textit{Wireless Communications Principles and Practice}, 2nd ed., Prentice Hall, 2001, pp. 69-138.

\bibitem{qpu_timing} 
``Operation and Timing" in \textit{D-Wave System Documentation}, 2021 [Online]. Available: https://docs.dwavesys.com/docs/latest/c\_qpu\_timing.html [Accessed 5 Apr. 2022].

\end{thebibliography}
\end{document}